\documentclass[a4paper,fleqn,usenatbib]{mnras}
\usepackage{natbib}
\usepackage[T1]{fontenc}
\usepackage{ae,aecompl}

\usepackage{graphicx}	
\usepackage{amsmath}	
\usepackage{amssymb}	

\title[UV emission from MS companions of AGB stars]{Ultraviolet emission
from main-sequence companions of AGB stars}

\author[R. Ortiz and M. Guerrero]{
Roberto Ortiz$^{1,2}$\thanks{E-mail: rortiz@usp.br}, 
Mart\'\i n A. Guerrero$^{1}$
\\
$^{1}$Instituto de Astrof\'\i sica de Andaluc\'\i a (IAA-CSIC), Glorieta de la Astronom\'\i a s/n, E-18008 Granada, Spain\\
$^{2}$Escola de Artes, Ci\^encias e Humanidades, USP, Av. Arlindo Bettio 1000, 03828-000 S\~ao Paulo, Brazil}

\date{Accepted XXX. Received YYY; in original form ZZZ}
\pubyear{2016}
\begin{document}
\label{firstpage}
\pagerange{\pageref{firstpage}--\pageref{lastpage}}
\maketitle

\begin{abstract}

Although the majority of known binary Asymptotic Giant Branch (AGB)
stars are symbiotic systems (i.e. with a WD as a secondary star),
main-sequence companions of AGB stars can be more numerous, even
though they are more difficult to find because the primary high
luminosity hampers the detection of the companion at visual wavelengths.
However, in the ultraviolet the flux emitted by a secondary with
$T_{\rm eff} > 5500 \sim 6000$ K may prevail over that of the primary,
and then it can be used to search for candidates to binary AGB stars.
In this work, theoretical atmosphere models are used to calculate the
UV excess in the \emph{GALEX} near- and far-UV bands due to a main-sequence
companion. After analysing a sample of confirmed binary AGB stars,
we propose as a criterium for binarity: (1) the detection of the AGB
star in the \emph{GALEX} far-UV band and/or (2) a \emph{GALEX} near-UV
observed-to-predicted flux ratio $>20$.
These criteria have been applied to a volume-limited sample of AGB stars
within 500 pc of the Sun; 34 out of the sample of 58 AGB stars
($\sim$60\%) fulfill them, implying to have a MS companion of spectral
type earlier than K0. The excess in the \emph{GALEX} near- and far-UV
bands cannot be attributed to a single temperature companion star, thus
suggesting that the UV emission of the secondary might be absorbed by
the extended atmosphere and circumstellar envelope of the primary or
that UV emission is produced in accretion flows.

\end{abstract}

\begin{keywords}
stars: {\sc agb} and post-{\sc agb} -- binaries: general -- circumstellar matter --
ultraviolet: stars
\end{keywords}

\section{Introduction}

Binarity has long been suggested as a mechanism to shape bipolar planetary 
nebulae (PNe) \citep[e.g.][]{CS93,Soker98}.
Observations have confirmed that a significant number of PNe
indeed have binary central stars \citep{DeMarco13}.
Since binarity precedes the formation of the PN, it is paramount 
to detect it along previous phases of stellar evolution before 
the PN formation, particularly at the Asymptotic Giant Branch
(AGB) phase.
An unbiased comparison among the binarity occurrence rates during the
main sequence, AGB and PN phases can help to reinforce a causality
relationship between binarity and the formation of aspherical PNe and
to assess the evolution of binary systems \citep[e.g.][]{Ivanova13,Staff16}

Binary or multiple systems including AGB stars have been often observed
as symbiotic systems. They are identified by their spectra, which includes
features characteristic of the red giant as well as emission lines arising
from the wind-driven atmosphere of the giant, which is ionized by the UV
photons of the secondary white dwarf (WD).  
The latest catalogue of symbiotic systems contains about two hundred 
objects, including confirmed and suspected objects \citep{Belczynski00}.

The discovery of hot companions of AGB stars is somewhat straightforward
where they compose symbiotic systems, but the detection of low- and 
intermediate-mass main-sequence (hereafter MS) companions is not simple.  
The detection of the secondary in direct images is difficult because
the high brightness contrast between them hampers the detection of
the secondary, except in the cases where the components are well resolved
\citep{Karovska93,Karovska97,Prieur02}.
Indeed, the recent advent of new generation adaptive optics systems
has allowed the detection of late-type, close companions of AGB stars
\citep{Beuzit08,Fusco14}, but this method still remains restricted to
a few near objects \citep{Kervella15}.
Other methods to detect binary AGB stars include asymmetries in their
circumstellar envelopes \citep{Mayer13}, proper-motion variations
\citep{Pourbaix03}, and the identification of features attributed
to the secondary in the visual spectrum \citep{CM95,Danilovich15}.

As a rule, the shorter the wavelength, the higher the relative contribution
of the hot component to the spectrum because the flux emitted by an AGB star
decreases abruptly beyond $\sim$2800~\AA.
Therefore, UV space observatories (\emph{FUSE}, \emph{GALEX}, \emph{HST})
have greatly increased the possibilities to detect MS companions of AGB
stars.  
\citet{Sahai08,Sahai11} have carried out a program in quest of binary 
AGB stars based on an imaging survey obtained by the \emph{GALEX} 
observatory \citep[{\it Galaxy Evolution Explorer,}][]{Martin05}.  
They selected 25 AGB stars showing M5 or later spectral type classified
during the mission as ``bright star'' and ``high-background'', with the
additional criterium that they should exhibit the ``multiplicity'' flag
in the catalogue.
These were considered as promising conditions to detect companions with
spectral type earlier than G0 in the two \emph{GALEX} photometric bands:
far-UV ($1340-1790$ \AA) and near-UV ($1770-2830$ \AA).
Indeed, UV counterparts were detected in most of the sources and for a
significant fraction of them the UV emission seemed to be in excess,
i.e.\ it probably results from a companion undetected at visual wavelengths.

These results have undoubtfully made a major contribution in increasing
the number of known binary AGB stars, but these searches have mostly been
biased towards the most promising candidates and/or been restricted to
symbiotic systems.
In this paper we address the problem of detecting MS companions of AGB 
stars using \emph{GALEX} UV data.
The issues explored in this study will pave the way for future unbiased
determinations of the occurrence of binarity among AGB stars.
The main scopes of the present paper are:
(1) to establish the detection limits of the \emph{GALEX} survey to
    detect MS companions of AGB binary stars (Sect.\ \ref{detect} and
    \ref{method});
(2) to derive criteria for selecting binary AGB star candidates, to be
    eventually confirmed by other techniques, e.g.\ radial velocity (RV) 
    studies 
    (Sect.\ \ref{results}); and 
(3) to assess the stellar properties of these companions, mainly their
    effective temperature (Sect.\ \ref{discussion}).

\section{The detectability of AGB companions with GALEX }
\label{detect}

The detection of the MS secondary companion of an AGB star 
depends on the contrast between its flux and that of the 
primary.
The detectability of a MS intermediate or early spectral type companion 
using UV photometry is feasible because, despite the high luminosity of 
the primary, its flux bluewards of 2800 \AA\ is usually negligible when 
compared to the flux emitted by the secondary.
For instance, \citet{Sahai08} estimated a secondary-to-primary \emph{GALEX} 
near-UV flux contrast ratio $\geq$10 for a MS secondary star with spectral
type earlier than G0 (or $T_{\rm eff} > 6000$ K).

Besides the flux contrast between the two components, the detectability
depends also on the distance to the system as it determines the UV flux.
This issue was extensively addressed by \citet{Bianchi07} for hot objects.
Assuming theoretical atmosphere models, they conclude that \emph{GALEX}
could detect all WD in the Galactic halo along their constant-luminosity
phase (i.e.\ WD's with $T > 50,000$ K and radii down to $0.04 R_{\odot}$) 
up to a distance of 20 kpc.
Concerning \emph{GALEX} detections of MS stars, \citet{Bianchi07}
focused on objects with $T_{\rm eff} > 18,000$ K, thus excluding
intermediate-mass stars.

\emph{GALEX} observation strategy was mainly organized into three
different modes, namely {\sc ais}, {\sc mis} and {\sc dis}, mostly
different by their varying exposure times of 100 s, 1500 s, and
30,000 s, respectively.
These modes refer to specific areas in the sky called {\it tiles.}
Most observations were made in the {\sc ais} mode to cover 
large areas in the sky.
For each entry the \emph{GALEX} catalogue\footnote{galex.stsci.edu}
gives, among other information, the magnitude and flux density
in the near- and far-UV bands \citep{Morrissey05}.
An analysis of sources registered in different mode tiles shows 
that there are significant differences in sensitivity among them: 
sources detected with a signal-to-noise ratio $(F/{\sigma}_F)_{\it NUV,FUV}$
of 5 have fluxes 10 $\mu$Jy $< F_{\it NUV,FUV} <$ 20 $\mu$Jy in {\sc ais}
tiles, going down to 5 $\mu$Jy $< F_{\it NUV,FUV} <$ 10 $\mu$Jy in {\sc mis}
tiles given their longer integration time.

In order to calculate the limiting distances of MS stars detectable by
\emph{GALEX}, we adopt the MS visual absolute magnitudes given by 
\citet{Cox02} and all the theoretical atmosphere models provided by
\citet{Lejeune97} for solar metallicity stars and spectral type later
than B2{\sc v}, i.e.\ for intermediate-mass stars.
The grid of \citet{Lejeune97} models is equally paced in $\log g$
by 0.5 dex intervals, but unequally in $T_{\rm eff}$.
In all cases the difference between the $\log g$ values given 
by \citet{Cox02} and the value of the adopted theoretical model 
never exceeded 0.2 dex. The relative differences in temperature were 
$(\Delta T_{\rm eff}/T_{\rm eff}) <$ 2.3\%.
Absolute (i.e. at a distance of 10 parsecs) flux densities in the $V$
band were calculated by convolving the theoretical spectrum with the
$V$ filter curve given by \citet{MA06} using equation:

\begin{equation}
F_X = \frac{\int_{0}^{\infty} F_{\rm \star}(\lambda)S_X(\lambda)d\lambda}
{\int_{0}^{\infty} S_X(\lambda) d\lambda},
\label{Fx}
\end{equation}

\noindent
where $X$ is the name of the photometric band ($V$ in this case),
$F_{\rm \star}(\lambda)$ is the theoretical flux density of the star, and 
$S_X(\lambda)$ is the filter response curve.
$F_{\rm \star}(\lambda)$ is ``scaled'' in order to match its flux density
at ${\lambda}_{\rm eff}=5500$ \AA\ with the value corresponding to its
absolute visual magnitude $M_V$ by \citet{Cox02}.
The visual zero-magnitude density flux used in this calibration is also
given by \citet{Cox02}:
$F_{\rm zero,V}=3.75 \times 10^{-9}$ erg s$^{-1}$cm$^{-2}$\AA$^{-1}$.
Finally, \emph{GALEX} near- and far-UV flux densities are calculated
from this scaled theoretical spectrum using the same Equation~\ref{Fx},
but for the near- and far-UV filter curves given by \citet{Morrissey05}.

\begin{figure}
\includegraphics[width=\columnwidth]{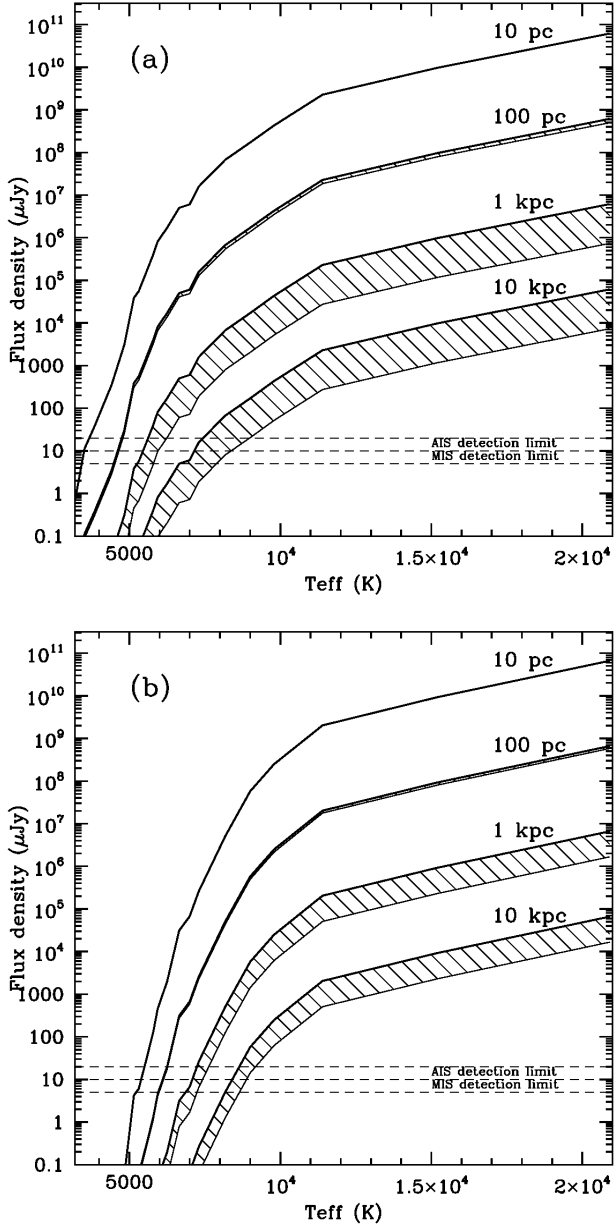}
\caption{
\emph{GALEX} near-UV {\bf (a)} and far-UV {\bf (b)} flux density
of MS stars at 10 pc, 100 pc, 1 kpc, and 10 kpc as a function of
effective temperature.
The thick lines represent cases for zero extinction, whereas the shaded
areas represent the flux densities affected by visual extinctions
$A_V=0$ mag for 10 pc, $0 < A_V \le 0.1$ mag for 100 pc, and
$0 < A_V \le 1.0$ mag for 1 and 10 kpc.
The horizontal dashed lines mark the range of {\sc ais} (i.e.
\emph{GALEX} observations with exposure time of $t=100$ s) and
{\sc mis} ($t=1500$ s) detection limits.}
\label{fig-detect}
\end{figure}

Figure \ref{fig-detect} shows the flux density of MS stars in the
near- and far-UV bands as a function of the stellar effective
temperature for distances in the range from 10 pc up to 10 kpc.
The interstellar extinction decreases the detection radius of the 
\emph{GALEX} survey.
To calculate this effect, we derive the extinction $A_{\it X}$
in the $X$ band ({\it X=FUV, NUV}) as follows:

\begin{equation}
A_X= R_X \times E(B-V),
\label{Anuv}
\end{equation}

\noindent
where
$E(B-V)$ is the colour excess in the direction of the star (extracted
from the \emph{GALEX} catalogue), and 
$R_{\it X}$ is the extinction coefficient in the {\it X} band.
The behaviour of the interstellar extinction coefficients was exhaustively
studied by \citet{Yuan13} who analysed thousands of stars detected by
\emph{GALEX} (and other surveys) in several directions in the Galaxy.
In the present study we adopt an UV extinction law that corresponds
to the average total-to-selective coefficient determined by \citet{Yuan13}:
$R_{\it NUV}=7.15$ and $R_{\it FUV}=4.63$.
Therefore the {\it extinction-corrected} ultraviolet flux density
$F_{\it X}$ can be calculated from Equation~\ref{Anuv} and the
corresponding extinction coefficients:

\begin{equation}
F_{\it X}=F^{\prime}_X \times 10^{+0.4 R_X E(B-V)},
\label{Fnuv}
\end{equation}

\noindent
where $F^{\prime}_X$ is the {\it observed} flux density in the near- and
far-UV bands.
Equation \ref{Fnuv} was applied to the theoretical flux densities shown
in Fig. \ref{fig-detect} to simulate the effect of the interstellar
extinction on near- and far-UV flux densities.
Because the extinction varies from source to source depending on their
distance $D$ and line of sight through the Galaxy, we simulate various
values of $A_V$ in Figure~\ref{fig-detect} from $A_V=0.1$ mag for
sources at $\simeq100$ pc to $A_V=1.0$ mag for $D\simeq1$ kpc
\citep[which corresponds to an average local extinction
of 1 mag~kpc$^{-1}$,][]{OL93,Marshall06,Froebrich10}.
The results are shown in the
same Fig. \ref{fig-detect}, where we applied $A_V=0.1$ mag to the 100 pc
and $A_V=1.0$ mag to the 1 and 10 kpc curves.
Assuming $R_V=A_V/E(B-V)=3.1$, the near-UV (far-UV) flux densities
would decrease by 19\% and 88\% (13\% and 75\%) for $A_V=0.1$ mag 
and $A_V=1.0$ mag, respectively.
Although a realistic estimation of the extinction should take into
account the distance and the Galactic coordinates of the object,
this simulation allows us to estimate the detection limits of MS
stars with \emph{GALEX}.  
MS stars with $T_{\rm eff} \gtrsim 5500$ K (spectral type G6 or earlier) 
could be detected in the near-UV band up to a distance of $\sim 1$ kpc 
in {\sc ais} tiles.
Contrary to naive expectations, only hotter stars with
$T_{\rm eff} \gtrsim 7000$ K (spectral type F2 or earlier,
i.e.\ mainly MS stars of spectral types O, B and A) could be
detected in the far-UV band due to its lower sensitivity.
The latter stars are less frequently companions of AGB stars because of
the steep slope of the initial-mass function.

\section{The method for estimating the near-ultraviolet excess of AGB stars}
\label{method}

The method used here for estimating the UV excess of AGB stars
has much in common with that proposed by \citet{Sahai08},
and has been partially described in Sect.~\ref{detect}.

Firstly, $B$ and $V$ magnitudes taken from {\it The Guide Star Catalogue,
GSC2.3} \citep{Lasker08} and/or \emph{HIPPARCOS} \citep{ESA97,Leeuwen07},
and $J$, $H$ and $K_s$ \citep[2MASS,][]{Skrutskie06} magnitudes are
corrected for interstellar extinction using the $E(B-V)$ values
listed in the \emph{GALEX} catalogue.
The standard values of $A_V=3.1 \times E(B-V)$ and $A_{\lambda}/A_V$ of
0.26, 0.15 and 0.09 for the $J$, $H$ and $K_s$ bands, respectively,
are adopted \citep{Koornneef83,Yuan13}.
Since $E(B-V) < 0.1$ mag in most cases, 
the assumption of an alternative reddening law would not change the 
results significantly.

Secondly, the extinction-corrected flux densities corresponding to
these five photometric bands are calculated using Equation~\ref{Fx}.
These are then fitted with the least-squared method to the theoretical
spectral library by \citet{Lejeune97} assuming [Fe/H]=0.
No photometric bands redwards of $K_s$ are considered in these fits
because AGB stars may have circumstellar dust envelopes that produce
an infrared excess in this spectral region.  
The average magnitude residuals considering the 5 bands is typically
less than 0.2 mag, which is considerably smaller than the amplitude
of variability of the stars.

Once the best-fit spectrum has been found, its flux density in the
\emph{GALEX} near-UV and far-UV bands is calculated using Equation~\ref{Fx}.
This latter step deserves a warning.
Since the AGB stellar spectrum in this spectral region is very steep
and the \emph{GALEX} photometric bands are broad, the flux densities
$F_{\it NUV}$ and $F_{\it FUV}$ obtained by convolving the theoretical
spectrum with the filter curve are dominated by the stellar contribution
on the longest wavelength side of the band.
As a result, the flux densities differ notably, about two orders of
magnitude, from the value of the flux at the effective wavelength
$F_{227\rm nm}$ and $F_{153\rm nm}$, respectively.

Finally, the ratios $Q_{\it NUV}$ and $Q_{\it FUV}$ of the 
extinction-corrected (using equation \ref{Fnuv}) \emph{GALEX} 
$F_{\it NUV}$ and $F_{\it FUV}$ to the predicted $F_{\it NUV}$ 
and $F_{\it FUV}$ are computed.

\section{Results}
\label{results}

In this Section we describe the results of the application of this
method to various samples of AGB stars. Firstly, we test it with the
same stars analysed by \citet[][Sect. \ref{results-sahai}]{Sahai08,Sahai11};
second, a sample of confirmed binary stars discovered using kinematical
criteria is studied (Sect. \ref{results-confirmed});
finally, the method is applied to a volume-limited sample of nearby
($< 0.5$ kpc) AGB stars (Sect. \ref{results-nearby}).

\begin{table*}
\caption{Basic data and best-fit parameters of the AGB stars studied
by \citet{Sahai08,Sahai11}. Distances are adopted from the \emph{HIPPARCOS}
catalogue. Single-value flux densities correspond to single-epoch
observations, whereas the intervals refer to the minimum and maximum
flux densities, obtained at multiple epochs.
All flux densities have been corrected for interstellar extinction.}
\label{table-sahai}
\begin{tabular}{lccccccccc}
\hline
Name & D & $T_{\rm eff}$ & $\log g$ & $E(B-V)$ &{\sc galex-}$F_{FUV}$ &
{\sc galex-}$F_{NUV}$ & {\sc predicted-} & $Q_{FUV}$ & $Q_{NUV}$ \\
 & (pc) & (K) & & & (mJy) & (mJy) & $F_{\it NUV}$ (mJy) & & \\
\hline
RW Boo & 293 & 2800 & $+0.28$ & 0.015 &0.0277 &0.430$-$0.519 & 0.0883 & $2.4 \times 10^{10}$ & 4.9$-$5.9 \\
AA Cam & 781  & 3350 & $+0.28$ & 0.044 & 0.0169 &0.294$-$0.374 & 0.0664 & $1.2 \times 10^7$ & 4.4$-$5.6 \\
V Eri  & 439 & 2800 & $-1.02$ & 0.040 & 0.0711 &0.182 & 0.0523 & $1.0 \times 10^{11}$ & 3.5 \\
R UMa  & 415  & 3500 & $+1.50$ & 0.025 &0.0456 &0.141 & 0.0448 & $1.3 \times 10^6$ & 3.2 \\
Y Gem  & 769 & 2800 & $+0.28$ & 0.051 &$27.13-414.8$&$1.90-11.30$ & 0.0491 & (0.4$-$6.4)$\times 10^{14}$ & 38.7$-$230.2 \\
$o$ Ceti & 92 & 3000 & $-0.29$ & 0.027 &$7.20-56.08$&$2.05-43.74$ & 0.434 & (0.7$-$5.5)$\times 10^{11}$ & 4.7$-$100.8 \\
\hline
\end{tabular}
\end{table*}

\subsection{Ultraviolet excess of M-type AGB stars (Sahai et al. 2008, 2011)}
\label{results-sahai}

\begin{figure*}
\includegraphics[width=16.0cm]{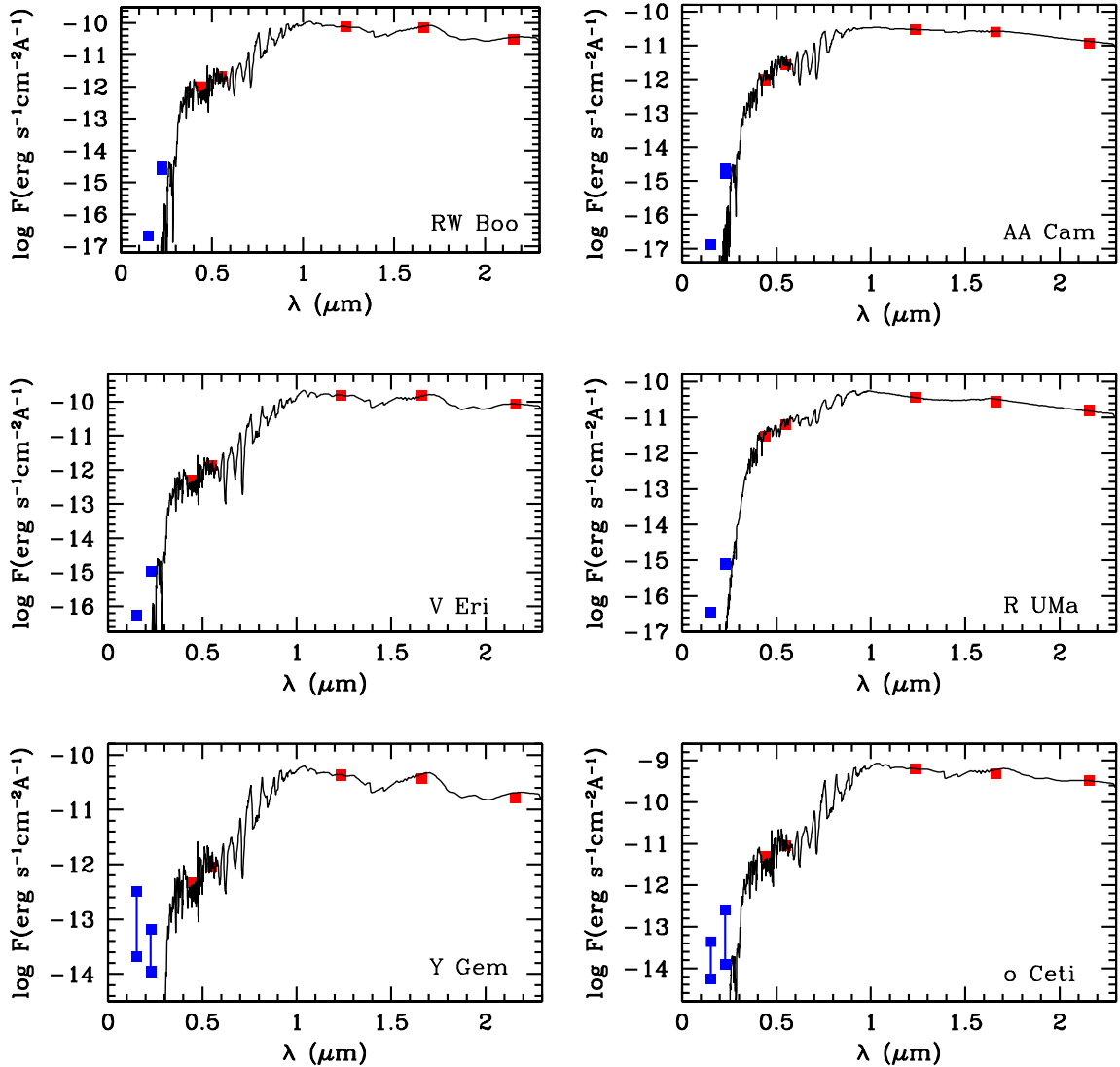}
\caption{
Flux densities and best-fit theoretical spectra of the stars studied
by \citet{Sahai08,Sahai11}.
The red squares correspond to the $B$, $V$ (GSC2 catalogue), $J$, $H$ and
$K_s$ (2MASS) photometric bands, and the blue squares to the \emph{GALEX}
near- and far-UV bands. All data have been corrected for interstellar
extinction.}
\label{6-stars}
\end{figure*}

\citet{Sahai08,Sahai11} analysed the UV excess of 6 M-type AGB
stars in quest of companions.  
We have applied our method to the sample in those references and show
the results in Table \ref{table-sahai}.
The range of $T_{\rm eff}$ observed for this small sample corresponds to
spectral types between M4$-$M8 \citep{Fluks94}, whereas the interval
determined by \citet{Sahai08,Sahai11} spans from M3$-$M9.
Table 2 of \citet{Lejeune97} lists $\log g$ as a function of the effective
temperature for a sequence of cool giants.
An interpolation of that set of values gives
$\log g=-0.78$ for $T_{\rm eff}=2800$~K, 
$\log g=-0.33$ for $T_{\rm eff}=3000$~K,
$\log g=+0.42$ for $T_{\rm eff}=3350$~K, and
$\log g=+0.71$ for $T_{\rm eff}=3500$~K.  
The differences between these values and those determined in Table
\ref{table-sahai} does not exceed 1.1 dex.
Figure \ref{6-stars} illustrates the photometric data and the
best-fit spectra.  
In the cases where multiple measurements of flux were taken at different
epochs, the maximum and minimum values are plotted.
The average residual between the best-fit model and the observed
{\it BVJHK}$_s$ magnitudes is smaller than 0.11 magnitude for the
6 stars studied.

All stars were detected both in the near- and far-UV bands.
In those cases when multiple measurements are available, the UV fluxes 
varied up to one order of magnitude.
Four stars (RW Boo, AA Cam, V Eri and R UMa) have $3 < Q_{\it NUV} < 6$.
Y Gem and $o$ Cet show much higher near-UV excess, up to $Q_{\it NUV} >100$.
Variability certainly accounts for at least a part of the near-UV
excess observed in these stars.
\citet{Celis86} monitored a sample of Mira- and SR-type variables in 
the {\it UBVRI} system and observed variations of over 3 magnitudes 
(a factor 16 in flux) in the $U$ band.

The predicted far-UV flux density is lower than $10^{-4}$ $\mu$Jy for
all the stars in the sample, which corresponds to an excess of
$Q_{\it FUV} > 10^6$, similar to the figure determined by \citet{Sahai08}.
The stars that show the larger near-UV excesses also show the larger
$Q_{\it FUV}$ values, even though the excess in the far-UV is several
order of magnitude larger than in the near-UV.
The far-UV excess observed in all the objects of the sample
indicates that they all must have an additional source of UV photons
such as an intermediate-mass MS star.
The high values of $Q_{\it NUV}$ reinforce this conclusion.

\begin{table*}
\caption{Basic data and best-fit parameters of AGB stars confirmed to be
binary \citep{Famaey09}.
All distances were determined from \emph{HIPPARCOS} parallaxes.
$Q_{\it NUV}$ ($Q_{\it FUV}$) represents the observed-to-predicted
$F_{\it NUV}$ ($F_{\it FUV}$) ratio.
The predicted $F_{\it FUV}$ of all the stars in this Table are lower
than $10^{-4} \mu$Jy and are not shown here.
When multiple-epoch \emph{GALEX} observations are available, flux density
minima and maxima are shown.
All flux densities have been corrected for interstellar extinction.}
\label{table-famaey}
\begin{tabular}{lccrcccccc}
\hline
Name & D & $T_{\rm eff}$ & $\log g$ & $E(B-V)$ &{\sc galex-}$F_{\it FUV}$ &
{\sc galex-}$F_{\it NUV}$ & {\sc predicted-} & $Q_{\it FUV}$ &
$Q_{\it NUV}$ \\
 & (pc) & (K) & & & ($\mu$Jy) & ($\mu$Jy) & $F_{\it NUV}$($\mu$Jy) & & \\
\hline
\sc{hip}50801  & 76 & 3750 & 0.00 & 0.013 &732.6 & 9535.9 & 253.0 & $1.1 \times 10^7$ & 37.7 \\
\sc{hip}58545 &251 & 3500 & 3.00 & 0.069 &105.8 & 1322.8$-$1483.8 & 127.8 & $2.6 \times 10^5$ & 10.5$-$11.6 \\
\sc{hip}60998 &178 & 3500 & 3.00 & 0.022 &... &7441.1 & 300.9 & ... & 24.7 \\
\sc{hip}62355 &490 & 3500 & 0.50 & 0.010 &665.6 &1403.9 & 7.0 & $3.1 \times 10^8$ & 200.1 \\
\sc{hip}72208 &279 & 3350 &$-0.29$ & 0.021 &5419.5$-$6155.0 &6833.6$-$10,818.5 & 290.4 & (8.7$-$9.9) $\times 10^8$ & 23.5$-$37.3 \\
\sc{hip}73199 &122 & 3500 & 0.00 & 0.023 &1991.3$-$2126.9 &7552.4$-$9110.4 & 114.8 & $1.0 \times 10^8$ & 65.8$-$79.3 \\
\sc{hip}74253 &532 & 3750 & 0.00 & 0.041 &23.6$-$38.3 &604.2$-$632.0 & 7.5 & (1.2$-$2.0)$\times 10^7$ & 80.6$-$84.3 \\
\sc{hip}81497 &115 & 3750 & 1.50 & 0.022 &100.4$-$150.5 &7626.8$-$8666.0 & 221.5 & (4.2$-$6.3) $\times 10^5$ & 34.4$-$39.1 \\
\sc{hip}84345 &117 & 3350 &$-0.29$ & 0.120 &21,917.3 &111,481.9 & 4270.3 & $2.4 \times 10^8$ & 26.1 \\
\sc{hip}88563 &275 & 3750 & 0.00 & 0.114 &38.3$-$43.0 &800.4$-$856.6 & 9.6 & (1.5$-$1.7)$\times 10^7$ & 83.4$-$89.2 \\
\sc{hip}97372 &377 & 3750 & 0.50 & 0.106 &97.2$-$112.6 &1820.6$-$2020.1 & 26.5 & (0.9$-$1.1)$\times 10^7$ & 68.7$-$76.2 \\
\sc{hip}110346 &243 & 3500 & 0.50 & 0.083 &91.1 &1799.6 & 26.2 & $1.2 \times 10^7$ & 68.7 \\
\hline
\end{tabular}
\end{table*}

\subsection{Ultraviolet emission of a sample of confirmed binary AGB stars}
\label{results-confirmed}

Radial velocity monitoring of spectroscopic binaries is a widely
used technique to derive their orbital parameters.
\citet{Famaey09} comment that only 1.1\% of the stars in the Ninth
Catalogue of Spectroscopic Binary Orbits \citep{Pourbaix04} are
M-type giants.
Indeed, most of the binary systems known to contain a red giant are
symbiotic, whereas systems with a non-degenerate companion are scarce.  
\citet{Famaey09} used a series of radial velocity measurements
obtained with the {\sc coravel} spectrovelocimeter \citep{Baranne79}
to determine the binarity status of a number of stars.
Binarity was flagged as {\it ORB} when the set of kinematic measurements
allowed the determination of the orbit, 
{\it ORB:} when the orbit was poorly determined,
{\it SB} when binarity was confirmed spectroscopically, but no orbit
could be computed from the data,
{\it SB?} when the binarity is only suspected, and 
{\it NON-SB} when binarity is discarded.

Our sample of confirmed binary AGB stars was extracted from the
work by \citet{Famaey09}, with the following main additional criteria:
(a) the star should be classified as {\it ORB}, {\it ORB:} or {\it SB};
(b) concerning variability, the star must be classified as
Mira (a {\it long-period variable} or LPV), SR- ({\it semi-regular})
or L-type ({\it slow irregular variable}). The former criterium is devised
to select confirmed AGB binaries and the latter intends to discard
M-stars belonging to the RGB phase. We also added some objects ({\sc hd}62898,
{\sc hd}108907, {\sc hd}130144, {\sc hd}150450, {\sc hd}156014 and {\sc hd}187372) that are associated
with X-ray \emph{ROSAT} sources, as suggested by \citet{Hunsch98}.
Stars with large X-ray offsets were not included because of their doubtful
association.
Finally, symbiotic systems, which do not represent the main scope of this
paper, have been also rejected.
Two stars ({\sc hd}16058 and {\sc hd}42995) were not surveyed by \emph{GALEX}.
Our final list, as shown in Table \ref{table-famaey}, contains 12
confirmed binary stars.
Their properties were obtained similarly as those of the sample
analysed in Sect.~\ref{results-sahai}.

Table \ref{table-famaey} shows that the discrepancy between the
\emph{GALEX} measured and predicted $F_{\it NUV}$ persists, but
on average $Q_{\it NUV}$ is higher than in the sample analysed in
Sect.\ \ref{results-sahai}.
Except in the case of {\sc hip}58545 ({\sc hd}104216 = FR Cam), $Q_{\it NUV}$
is generally comparable to the highest values in Sahai's sample,
(Y Gem and $o$ Ceti), i.e. $10 < Q_{\it NUV} < 200$.
Two general properties can be derived from this sample of confirmed AGB
stars: $Q_{\it NUV} > 20$ (except in the case of FR Cam) and
the detection of a far-UV counterpart (except {\sc hip}60998 = {\sc hd}108907 = 4 Dra).

\subsection{Ultraviolet emission of nearby AGB stars}
\label{results-nearby}

After having tested the method proposed by \citet{Sahai08} to
investigate the UV excess associated with confirmed binary AGB stars,
we extend it to a volume-limited sample of AGB stars.

There are different lists of nearby AGB stars in the literature that
could be used to assamble a volume-limited sample of AGB stars
\citep{Sivagnanam88,Groenewegen92,Jura94,OM96}.
However, all these ``old'' lists suffer from large uncertainties in the
determination of the stellar distances \citep{vL90,LOE95,Glass95,GW96}.
This limitation has been overcome with the advent of \emph{HIPPARCOS},
which allowed the determination of visual photometry, position, proper
motion and annual parallax with accuracy starting at
${\sigma}_{\pi}=0.6$ {\it mas} (milli-arcsec) for magnitudes $V = 5-6$ up to
${\sigma}_{\pi}=2.5-3.5$ {\it mas} near the limiting magnitude $V = 12.4$ of
the survey \citep{Perryman97}.
Thus, since ${\sigma}_{\pi}$ increases with the apparent magnitude
and $\pi$ decreases with the distance, the relative uncertainty
${\sigma}_{\pi}/\pi$ for a star generally increases with its distance.

Red giant stars were selected from the list of \emph{HIPPARCOS} sources
classified by \citet{Ita10} as M-, C-, S- and OH/IR stars.
In order to remove RGB stars from that list, we selected only
stars with luminosity higher than the tip of the RGB
\citep[$L > 3 \times 10^3 L_{\odot}$,][]{SC97} obtained by
\citet{McDonald12}.
About 95\% of the AGB stars within 500 pc selected according to this
criterium have apparent visual magnitudes $V < 9.0$, and consequently
${\sigma}_{\pi}/\pi < 0.5$.
Beyond this distance, the distance uncertainties based on \emph{HIPPARCOS}
data are larger than those reported by other formerly proposed distance
scales.
Therefore, the volume-limited sample of AGB stars in the solar neighbourhood
studied in this paper is limited to a distance of 500 pc.

After applying the above criteria, the sample is reduced to 90 AGB stars,
from which we additionally discarded those that 
(1) are classified in the literature as symbiotic (1 object, R Aqr =
{\sc hip}117054),
(2) were not observed by \emph{GALEX} (31 objects), and 
(3) were not {\it detected} by \emph{GALEX} (5 objects).  
Our final sample consists of 53 AGB stars of various types within 500
pc, with \emph{GALEX} counterparts and $L/L_{\odot}> 3 \times 10^3$
(Table~\ref{table-Ita}).
According to Figure~\ref{fig-detect}, the limiting temperatures for the
detection of MS stars at a distance of 500 pc in \emph{GALEX} data would
be $T_{\rm eff} \gtrsim 5000$ K for the near-UV and $T_{\rm eff} \gtrsim
6500$ K, for the far-UV.

\begin{table*}
\caption{Basic data and best-fit parameters of nearby (D$< 500$ parsecs)
AGB stars with \emph{HIPPARCOS} distances, selected from the list
of \citet{Ita10}.
Flux density intervals refer to minimum and maximum values obtained at
multiple epochs.
Near- and far-UV flux densities have been corrected
for interstellar extinction.}
\label{table-Ita}
\begin{tabular}{lccrcccccc}
\hline
Name & D & $T_{\rm eff}$ & $\log g$ & $E(B-V)$ &{\sc galex-}$F_{\it FUV}$ &
{\sc galex-}$F_{\it NUV}$ & {\sc predicted-} & $Q_{\it FUV}$ &
$Q_{\it NUV}$ \\
 & (pc) & (K) & & & ($\mu$Jy) & ($\mu$Jy) & $F_{\it NUV}$($\mu$Jy) & &  \\
\hline
\multicolumn{10}{c}{C-stars:} \\
{\sc hip}5914 & 385 & 3500 &  0.00 & 0.110 &     ...     &  50.1$-$155.5 & 15.1      & ... & 3.3$-$10.3 \\
{\sc hip}14930 & 322 & 3500 &  0.00 & 0.022 & 28.4$-$36.1 & 299.9$-$848.0 & 27.7      &(5.7$-$7.3)$\times 10^6$ & 10.8$-$30.6 \\
{\sc hip}43811 & 342 & 3350 &$-0.29$& 0.023 & 7.8         & 8.4           & 112.4     &$3.2 \times 10^6$ & 0.1 \\
{\sc hip}52009 & 208 & 3500 &  0.00 & 0.051 & 6.1         & 221.6         & 61.9 & $5.5 \times 10^5$ & 3.6 \\
{\sc hip}52577 & 380 & 3500 &  0.00 & 0.016 & 7.0$-$10.4  & 93.8$-$260.5  & 21.0 & (1.9$-$2.8)$\times 10^6$ & 4.5$-$12.4 \\
{\sc hip}62223 & 321 & 3350 &$-0.29$& 0.023 &     ...     & 7.4$-$9.3     & 191.8 & ... & 0.04$-$0.05 \\
{\sc hip}63152 & 431 & 3350 &$-0.29$& 0.033 &     ...     & 2.8           & 65.8 & ... & 0.04 \\
{\sc hip}91703 & 358 & 3200 &  0.28 & 0.224 &     ...     & 78.3          & 108.9 & ... & 0.7 \\
{\sc hip}95154 & 386 & 3500 &  0.00 & 0.094 &     ...     & 49.9          & 20.9 & ... & 2.4 \\
{\sc hip}117245 & 275 & 3500 &  0.00 & 0.056 & 10.7        & 331.8         & 43.2 & $1.4 \times 10^6$ & 7.7 \\
\multicolumn{10}{c}{M-stars:} \\
{\sc hip}1236 & 337 & 3000 &  0.60 & 0.015 &     ...     & 30.6$-$37.6   & 119.5 & ... & 0.3 \\
{\sc hip}2086 & 483 & 3500 &  3.00 & 0.023 & 40.9        & 969.3         & 114.0 & $1.1 \times 10^5$ & 8.5 \\
{\sc hip}9234 & 174 & 2500 &  0.28 & 0.117 &     ...     & 137.2$-$150.7 & 191.8 & ... & 0.7$-$0.8 \\
{\sc hip}13384 & 368 & 3350 &  0.28 & 0.048 & 33.2$-$34.0 & 524.4$-$665.2 & 153.9 & $1.0 \times 10^7$ & 3.4$-$4.3 \\
{\sc hip}17881 & 260 & 3200 &  0.28 & 0.151 & 47.7        & 888.9$-$1665.4& 336.3 & $2.6 \times 10^7$ & 2.6$-$5.0 \\
{\sc hip}20075 & 463 & 3500 &  1.50 & 0.046 & 96.1        & 1863.1        & 76.6 & $1.6 \times 10^6$ & 24.3 \\
{\sc hip}26169 & 333 & 3500 &  3.00 & 0.122 &125.6$-$159.8&2846.3$-$3181.3& 256.0 & (1.6$-$2.0)$\times 10^5$ & 11.1$-$12.4 \\
{\sc hip}28041 & 437 & 2800 &  0.60 & 1.112 &     ...     & 62931.8       & 381.5 & ... & 164.9 \\
{\sc hip}28874 & 203 & 3200 &  0.28 & 0.038 & 65.9        & 482.2         & 298.3 & $4.1 \times 10^7$ & 1.6 \\
{\sc hip}38792 & 341 & 4500 &  3.00 & 1.022 &     ...  &$8.6 \times 10^5$ &$1.7 \times 10^4$ & ... & 49.6 \\
{\sc hip}41028 & 450 & 2800 &  0.60 & 0.038 & 13.1$-$16.1 & 92.9$-$94.4   & 62.2 & (1.6$-$2.0)$\times 10^{10}$ & 1.5 \\
{\sc hip}53809 & 261 & 2500 &$-1.02$& 0.050 &     ...     &51.8$-$53.3    & 66.6 & ... & 0.8 \\
{\sc hip}53940 & 448 & 3500 &  2.00 & 0.138 & 97.3        & 1944.6        & 105.8 & $7.9 \times 10^5$ & 18.4 \\
{\sc hip}57607 & 242 & 3200 &$-0.29$& 0.113 &     ...     & 893.2$-$894.5 & 317.7 & ... & 2.8 \\
{\sc hip}58225 & 356 & 3200 &  0.60 & 0.028 & 4.0         & 248.3$-$288.0 & 165.8 & $4.4 \times 10^6$ & 1.5$-$1.7 \\
{\sc hip}62611 & 405 & 3500 &  1.50 & 0.060 & 55.8        & 692.0         & 51.9 & $1.4 \times 10^6$ & 13.3 \\
{\sc hip}64569 & 143 & 3000 & -0.70 & 0.034 &     ...     & 86.9          & 208.9 & ... & 0.4 \\
{\sc hip}72208 &279 & 3350 &$-0.29$ & 0.021 &5419.5$-$6155.0 &6833.6$-$10,818.5 & 290.4 & (8.7$-$9.9)$\times 10^8$ & 23.5$-$37.3 \\
{\sc hip}73213 & 403 & 3200 &  0.60 & 0.076 &     ...     & 335.9         & 248.8 & ... & 1.4 \\
{\sc hip}76075 & 305 & 3500 &  2.00 & 0.189 & 90.6        &1518.3$-$1646.2& 140.2 & $5.5 \times 10^5$ & 10.8$-$11.7 \\
{\sc hip}80488 & 235 & 3000 &$-0.29$& 0.059 &     ...     & 58.2          & 102.7 & ... & 0.6 \\
{\sc hip}80704 & 109 & 3200 &  0.60 & 0.009 & 46.3        & 3097.7        & 1878.4 & $4.6 \times 10^6$ & 1.6 \\
{\sc hip}98438 & 312 & 3750 &  3.00 & 0.368 &     ...     & 26,058.7      & 1093.0 & ... & 23.8 \\
{\sc hip}98608 & 146 & 3350 &  0.87 & 0.048 & 395.8       & 4423.3        & 927.5 & $2.0 \times 10^7$ & 4.8 \\
{\sc hip}99082 & 197 & 2500 &$-0.70$& 0.109 &     ...     & 84.8$-$101.2  & 126.1 & ... & 0.7$-$0.8 \\
{\sc hip}99920 & 318 & 3500 &  2.00 & 0.072 & 147.3       & 7618.4        & 117.5 & $1.1 \times 10^6$ & 64.8 \\
{\sc hip}100935 & 211 & 3000 &$-0.70$& 0.071 &     ...     & 154.4$-$492.0 & 216.1 & ... & 0.7$-$2.3 \\
{\sc hip}102978 & 258 & 4000 &  0.50 & 0.083 &     ...     &9676.3$-$9958.0& 181.4 & ... & 53.3$-$54.9 \\
{\sc hip}104451 & 188 & 3200 &  0.28 & 0.523 &     ...     &6038.5$-$25519.0& 1405.7 & ... & 4.3$-$18.2 \\
{\sc hip}106642 & 175 & 3500 &  2.50 & 0.556 &     ...     &55,342.9       & 1214.7 & ... & 45.6 \\
{\sc hip}110428 & 250 & 3000 &$-0.29$& 0.157 &     ...     & 101.1         & 90.2 & ... & 1.1 \\
{\sc hip}110736 & 446 & 2500 &$-1.02$& 0.012 &     ...     & 33.4$-$731.7  & 15.7 & ... & 2.1$-$46.8 \\
{\sc hip}114917 & 472 & 3000 &  0.60 & 0.035 &     ...     & 121.4         & 156.1 & ... & 0.8 \\
\multicolumn{10}{c}{OH/IR stars:} \\
{\sc hip}47886 & 347 & 2800 &$-0.70$& 0.015 &     2.5     &18.5$-$65.5    & 31.6 & $6.0 \times 10^9$ & 0.6$-$2.1 \\
\multicolumn{10}{c}{S-stars:} \\
{\sc hip}1728 & 270 & 3350 &  0.60 & 0.024 &    24.1     & 1488.4        & 464.1 & $2.4 \times 10^6$ & 3.2 \\
{\sc hip}1901 & 386 & 2500 &$-1.02$& 0.091 &     ...     &  50.1         & 12.5  & ... & 4.0 \\
{\sc hip}10687 & 448 & 3000 &  0.28 & 0.091 &     ...     & 13.9$-$27.3   & 88.6 & ... & 0.2$-$0.3 \\
{\sc hip}19853 & 450 & 3500 &  1.00 & 0.405 &     ...     &1210.3$-$1416.0& 46.2 & ... & 26.2$-$30.6 \\
{\sc hip}22667 & 200 & 3750 &  3.00 & 0.505 &  13,420.7   &83,351.5       & 2236.7 & $1.9 \times 10^6$ & 37.3 \\
{\sc hip}36288 & 253 & 3000 &  0.60 & 0.094 &     9.0     & 301.3$-$440.4 & 273.7 & $1.4 \times 10^8$ & 1.1$-$1.6 \\
{\sc hip}45058 & 143 & 2800 &  0.28 & 0.021 &     8.0     & 393.9         & 717.5 & $8.5 \times 10^8$ & 0.5 \\
{\sc hip}77619 & 293 & 3200 &$-0.70$& 0.021 & 39.9$-$47.8 & 341.1$-$412.0 & 111.0 & $6.7 \times 10^7$ & 3.1$-$3.7 \\
{\sc hip}113131 & 418 & 3500 &  3.00 & 0.091 &    42.4     & 1113.6        & 158.4 & $8.5 \times 10^4$ & 7.0 \\
\hline
\end{tabular}
\end{table*}

\begin{figure}
\includegraphics[width=8.0cm]{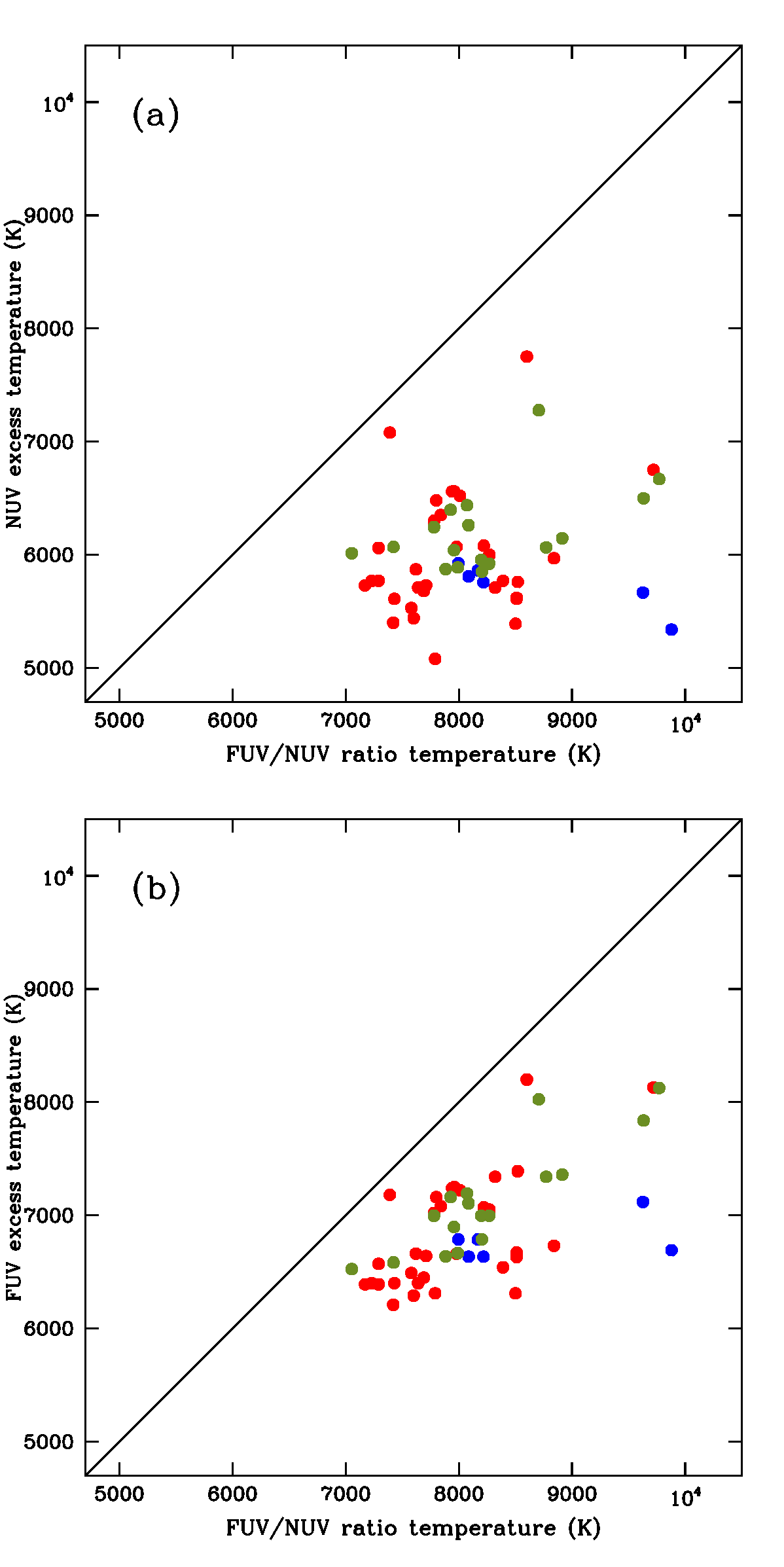}
\caption{
Temperature of the companion obtained from the $F_{\it FUV}/F_{\it NUV}$
excess ratio {\it versus} the temperature obtained from the near-UV
(\emph{top panel}) and the far-UV (\emph{bottom panel}) excess.
The different samples of AGB stars are shown using symbols with blue
(\citet[][Table \ref{table-sahai}]{Sahai08,Sahai11}), green
(\citet[][Table \ref{table-famaey}]{Famaey09}), and red
(distance-limited AGB stars, Table \ref{table-Ita}) points.  
When multiple UV observations are available, {\rm only} the temperatures
corresponding to the maximum and minimum values of the fluxes are shown.}
\label{fig-txt}
\end{figure}

The near-UV excess of the sample, expressed as $Q_{\it NUV}$, spans over
a wide range, from 0.1 to $\simeq$165.
Over 70\% of the stars show low near-UV excess, $Q_{\it NUV}<10$, and
only 12 stars (23\%) show near-UV excess similar to confirmed binary
AGB stars, i.e.\ $Q_{\it NUV}>20$.
Another significant statistical differences between this sample and that of 
confirmed AGB binaries are that only 50\% of the volume-limited sample (27
out of 53 stars) have a far-UV counterpart, while $Q_{\it FUV}$ spans over a
wider range of values, from $1.1\times10^5$ to $2.0\times10^{10}$.

\subsection{Temperature of the companions of AGB stars}
\label{temperature}

The UV excess observed in AGB stars can be used to assess the properties
of their putative companions, such as the temperature.
Assuming that the secondary component is a MS star, there is a close
relationship between its intrinsic flux and the effective temperature.
In this Section we obtain the temperature of the companion star
using two methods:
(1) the near-UV and the far-UV flux in excess of the AGB star
and (2) the near-to-far UV excess flux ratio.

Once the distance to the the star is known, the
UV flux density that would be emitted by a MS companion star is
calculated using the method described in Sect. \ref{method}.
Thus, the secondary's flux density is calculated as the difference
between the observed \emph{GALEX} and the predicted (theoretical)
flux density, listed in Tables \ref{table-sahai}, \ref{table-famaey}
and \ref{table-Ita}.
The temperature of the secondary star is given by the intersection
of the horizontal line corresponding to the source's UV excess and
the flux-density curve corresponding to the distance of the star in
the ${\rm T}_{\rm eff}$ {\it versus} $F_{\it NUV,FUV}$ diagram (Fig.
\ref{fig-detect}). The method is carried out separately in the near-
and far-UV \emph{GALEX} bands.

The second method is based on the $F_{\it FUV}/F_{\it NUV}$ excess ratio.
This quantity is compared with the prediction of a theoretical
model of MS stars, as discussed in Sect. \ref{detect}, being
thus similar to the temperature commonly determined from colour
indices.

Table \ref{table-dobles} lists the temperature of the AGB companions
obtained with the two methods described above.
In general, we observe the following inequality:
$T_{\it NUV} < T_{\it FUV} < T_{\it FUV/NUV}$ (see also Fig. \ref{fig-txt}).
This discrepancy suggests that the UV excess cannot be ascribed to a
single stellar spectrum.

\begin{figure}
\includegraphics[width=8.0cm]{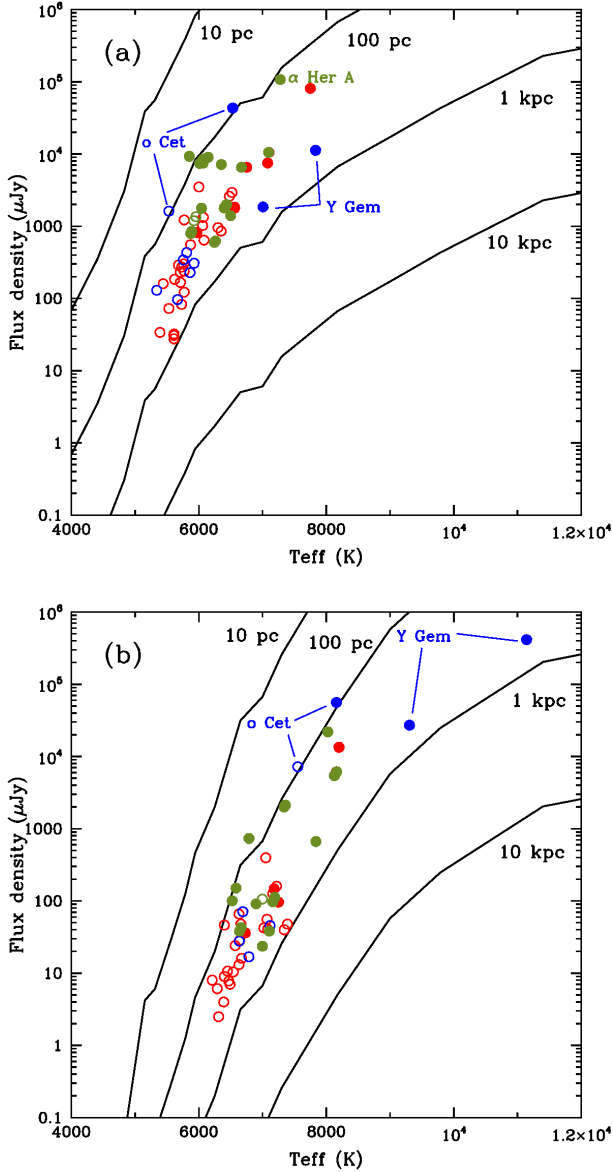}
\caption{
Effective temperature {\it versus} the near-UV {\it (a)} and far-UV
{\it (b)} {\sc galex} flux densities of the three samples studied in
this paper.
All fluxes have been corrected for interstellar extinction using
equation \ref{Fnuv}.
Colour code refers to different samples as in Figure~\ref{fig-txt}.
Filled symbols represent sources with $Q_{\it NUV} > 20$ and empty
symbols $Q_{\it NUV} < 20$.
In the case where multiple UV observations are available, the maximum
and minimum values of the fluxes are represented.}
\label{fig-temp}
\end{figure}

Figure~\ref{fig-temp} shows the temperature of the secondary members of
the three samples obtained with the first method, using the near-
and far-UV excesses. The reduced number
of stars closer than 100 parsecs (only $o$ Ceti and {\sc hd}89758 = $\mu$ UMa)
is a natural limitation caused by the relative low density of AGB stars
in the solar neighbourhood.
On the other hand, the relative larger number of objects near the
distance upper limit of 500 pc results from the fact that, assuming
a constant density of stars, the number of objects at a distance $D$
increases with $D^2$.
On average,
temperatures obtained from the far-UV excess flux are higher than
those determined from the near-UV band. There is also a selective
distribution of temperatures when the three distinct samples are compared,
but this is mostly due to selection effects.
The sample studied by \citet{Sahai08,Sahai11} and \citet{Famaey09}
are biased towards UV bright objects, whilst that described in
Sect.~\ref{results-nearby} is a distance-limited sample.

\citet{Sahai08,Sahai11}
used the second method to determine the temperature of the secondary,
i.e. based on the near-to-far UV flux density ratio and atmosphere models
by \citet{CK03}.
Table \ref{table-dobles} shows a good agreement between their results
and ours.
For instance, Y Gem has one of the highest $F_{\it FUV}/F_{\it NUV}$
ratio and \citet{Sahai11} argue that its companion might have a
blackbody temperature as high as 38,000 K, which is beyond the spectral
range considered in our study, restricted to intermediate-mass MS stars.
Our results, obtained with the same method, confirms this conclusion.
We note that some UV sources are associated with X-ray sources, but
no clear relationship seems to exist among them, neither with the
amount of far-UV excess, and/or the temperature of the secondary.

\section{Discussion}
\label{discussion}

\subsection{The effect of variability}
\label{variability}

One of the major concerns when fitting photometric or spectroscopic
data of AGB stars is their variability. This effect acts differently
along the spectrum, changing its shape as pulsation changes the stellar
temperature. As commented in Sect.
\ref{results-sahai}, Mira- and SR-type variables often show a peak-to-peak
variation of $\sim 3$ magnitudes (a factor $\sim 16$ in flux) in the $U$ band
\citep{Celis86}. The majority of the {\sc galex} measurements
were taken only once or twice, which is clearly insufficient to sample
the whole range of the UV fluxes along the cycle.
Therefore, in the cases where one or just a few multiple photometric
measurements of a given source were taken, the actual amplitude of
variability might be larger than the measured interval of fluxes.

The majority of the confirmed binaries listed in Table~\ref{table-famaey}
show a variation in flux of 10-20\%, whereas other systems, like
$o$ Ceti and Y Gem for example, show relative flux variations over one 
order of magnitude.
This suggests that the variability of the primary might affect the UV
flux emitted by its companion.
This effect is more evident in the far-UV band, where the predicted flux
of the primary is over 5 orders of magnitude smaller than the values
detected by {\sc galex}. Therefore, besides the intrinsic variability
of the primary's atmosphere, other effects might play a role in
the large UV amplitude observed in some binary AGB
stars.
For instance, it is well known that some AGB stars may exhibit large
circumstellar dust envelopes (CDE) that can extend up to 1000 
stellar radii.
The extended atmosphere of the AGB star and its CDE may block part of
the UV radiation emitted by the companion.
An accurate estimate of this effect depends on various factors, most
of them poorly known, such as the orbital parameters of the system,
including the inclination angle, the extinction law of the dust (which
depends on the chemical composition), the density law, and the opacity
of the CDE.
Figure \ref{fig-temp} shows that the variation of 1.3 dex in the near-UV
flux of $o$ Ceti causes a change of 1000~K in the determination of its 
temperature.
The effect on the far-UV band is even more important: the variation of
1.2 dex in flux of Y Gem modifies its temperature by 1800~K.

Another possiblity would be the UV emission produced by an accretion 
flow onto the secondary or an accretion disk around it.
This would produce stochastic or orbital-locked variations in the UV
flux \citep{Setal15}.

\begin{table*}
\caption{Confirmed and suspected binary AGB stars studied in this
paper. Temperatures were calculated according to the methods proposed in
Sect. \ref{temperature}. The references in the Table are: 
$a$=\citet{JAA93},$b$=\citet{SL87},$c$=\citet{Beck10}, $d$=\citet{Cox12},
$e$=\citet{AJ88},$f$=\citet{Cotton10},$g$=\citet{Fran07},$h$=\citet{GW96},
$i$=\citet{Pourbaix03},$j$=\citet{Famaey09},$k$=\citet{Nhung15},
$l$=\citet{Sanchez15},$m$=\citet{Mason99},$n$=\citet{Horch11},
$o$=\citet{Mason01},$p$=\citet{Sahai08},$q$=\citet{Sahai11},
$r$=\citet{Hunsch98},$s$=\citet{TR93}.}
\label{table-dobles}
\begin{tabular}{lccccccl}
\hline
Name & HIPPARCOS & $T_{\it NUV}$ & $T_{\it FUV}$ & $T_{\it NUV/FUV}$ & Far-UV & $Q_{\it NUV}$ & Comments\\
     & &(K)          & (K)          & (K)             &    (Y/N)    & &  \\
\hline
\multicolumn{8}{c}{{\it Table \ref{table-sahai}:}} \\
RW\,Boo  & {\sc hip}71802 & 5760$-$5810 &     6640      & 8090$-$8220 & Y & 4.9$-$5.9 &  companion: $T_{\rm eff}=8200$ K and $L=18 L_{\odot}^{p}$\\
AA\,Cam  & {\sc hip}35045 & 5860$-$5930 &     6790      & 8000$-$8170 & Y & 4.4$-$5.6 & companion: $T_{\rm eff}=8200$ K$^{p}$\\
V\,Eri   & {\sc hip}19004 &    5340    &     6690      & 9880 & Y & 3.5 &  companion: $T_{\rm eff}=10,000$ K$^{p}$\\
R\,UMa   & {\sc hip}52546 &    5670    &     7120      & 9630 & Y & 3.2 & companion: $T_{\rm eff}=9200$ K$^{p}$\\
Y\,Gem   & {\sc hip}37438 & 7010$-$7830 & 9310$-$11,140 & $>20,900$ & Y & 38.7$-$230.2 & companion: $T_{\rm eff}=17,000-38,000$ K$^{q}$\\
$o$\,Cet & {\sc hip}10826 & 5530$-$6530 & 7550$-$8160   & $>20,900$ & Y & 4.7$-$100.8 & symbiotic, WD secondary$^o$ \\
\hline
\multicolumn{8}{c}{{\it Table \ref{table-famaey}:}} \\
$\mu$\,UMa  & {\sc hip}50801 & 5850 & 6790 & 8200 & Y & 37.7 &  spectroscopic binary$^{j}$ \\
FR\,Cam & {\sc hip}58545 & 5920$-$5960 & 7000 & 8200$-$8270 & Y & 10.5$-$11.6 & spectroscopic binary$^{j}$ \\
4\,Dra & {\sc hip}60998 & 6350 & $\dots$ & $\dots$ & N & 24.7 & hot compact companion${^j}$\\
BY\,CVn & {\sc hip}62355 & 6500  & 7840 & 9630 & Y & 200.1 & spectroscopic binary$^{j}$ \\
EK\,Boo & {\sc hip}72208 & 6670$-$7090 & 8130$-$8160 & 9770$-$14,870 & Y & 23.5$-$37.3 & X-ray source$^{r}$ \\
RR\,UMi & {\sc hip}73199 & 6070$-$6150 & 7340$-$7360 & 8770$-$8910 & Y & 65.8$-$79.3 &  X-ray source$^{r}$ \\
FF\,Boo & {\sc hip}74253 & 6250$-$6260 & 7000$-$7110 & 7780$-$8090 & Y & 80.6$-$84.3 & spectroscopic binary$^{j}$ \\
42\,Her & {\sc hip}81497 & 6010$-$6070 & 6520$-$6580 & 7053$-$7420 & Y & 34.4$-$39.1 & X-ray source$^{r}$ \\
$\alpha$\,Her\,A & {\sc hip}84345 & 7280 & 8020 & 8710 & Y & 26.1 & $\alpha$\,Her\,B (G5{\sc iii}+A9{\sc iv-v}) at 4.7\arcsec $^s$\\
V980\,Her & {\sc hip}88563 & 5890 & 6670 & 7990 & Y & 83.4$-$89.2 & non-symbiotic$^{j}$\\
HR7547 & {\sc hip}97372 & 6400$-$6440 & 7160$-$7190 & 7930$-$8070 & Y & 68.7$-$76.2 & X-ray source$^{r}$ \\
PT\,Peg & {\sc hip}110346 & 6040 & 6900 & 7960 & Y & 68.7 & non-symbiotic$^{j}$ \\
\hline
\multicolumn{8}{c}{{\it Table \ref{table-Ita}:}} \\
T\,Cet & {\sc hip}1728 & 6060 & 6570 & 7290 & Y & 3.2 & upper limit on UV excess$^{a,b}$\\
{\sc hd}2268 & {\sc hip}2086 & 6350 & 7080 & 7840 & Y & 8.5 & \\
RR\,Eri & {\sc hip}13384 &5910$-$5980 & 6820$-$6830 & $>20,900$ & Y & 3.4$-$4.3 & \\
TW\,Hor & {\sc hip}14930 & 5730$-$5970 & 6640$-$6730 &7710$-$8840 & Y & 10.8$-$30.6 & binary$^{c}$, with no obvious bow shock$^{d}$ \\
SS\,Cep & {\sc hip}17881 & 5870$-$6070 & 6660 & 7620$-$7980 & Y & 2.6$-$5.0 & \\
V1139\,Tau & {\sc hip}19853 & 6380$-$6440 & ... & ... & N & 26.2$-$30.6 & undetected \emph{HIPPARCOS} binary$^{m}$ \\
GZ\,Eri & {\sc hip}20075 & 6560 & 7250 & 7960 & Y & 24.3 & \\
$o$\,Ori & {\sc hip}22667 & 7750 & 8200 & 8600 & Y & 37.3 & spec. binary with WD companion$^{e}$ \\
WX\,Men & {\sc hip}26169 &6480$-$6520 & 7160$-$7220 & 7800$-$8010 & Y & 11.1$-$12.4 & unresolved \emph{HIPPARCOS} ``Problem star''$^{m}$\\
U\,Ori & {\sc hip}28041 & 8690 & ... & ... & N & 164.9 & jet observed at 43 GHz SiO $J=1-0$ $^{f}$ \\
S\,Lep & {\sc hip}28874 & 5620 & 6630 & 8510 & Y & 1.6 & binary candidate$^{g}$\\
Y\,Lyn & {\sc hip}36288 &5610$-$5710 & 6400 &7430$-$7640 & Y & 1.1$-$1.6 & binary$^{d,h,i}$\\
PX\,Pup & {\sc hip}38792 & 8530 & ... & ... & N & 49.6 & \\
Z\,Cnc & {\sc hip}41028 &5610 &6630$-$6670 &8510$-$8640 & Y & 1.5 & \\
X\,Cnc & {\sc hip}43811 & 4980 & 6470 & 14,030 & Y & 0.1 & suspected binary$^{j}$\\
RS\,Cnc & {\sc hip}45058 & 5400 & 6210 & 7420 & Y & 0.5 & asymmetric CSE$^{k}$\\
R\,LMi & {\sc hip}47886 & 5080$-$5390 & 6310 & 7790$-$8500 & Y & 0.6$-$2.1 & jet observed at 43 GHz SiO $J=1-0$ $^{f}$\\
U\,Hya & {\sc hip}52009 & 5440 & 6290 & 7600 & Y & 3.6 & UV emission from a detached shell$^{l}$\\
VY\,UMa & {\sc hip}52577 &5530$-$5770 &6490$-$6540 & 7580$-$8390 & Y & 4.5$-$12.4 & binary$^{c,d}$\\
V361\,Vel & {\sc hip}53940 & 6560 & 7240 & 7940 & Y & 18.4 & \\
Z\,UMa & {\sc hip}58225 &5730$-$5770 & 6390 & 7170$-$7290 & Y & 1.5$-$1.7 & \\
SV\,Crv & {\sc hip}62611 & 6080 & 7070 & 8220 & Y & 13.3 & unresolved \emph{HIPPARCOS} ``Problem star''$^{m}$\\
EK\,Boo & {\sc hip}72208 & 6750$-$7100 & 8130$-$8160 &9720$-$12040 & Y & 23.5$-$37.3 & double star, separation 0.2\arcsec $^{o}$\\
GG\,Lib & {\sc hip}76075 & 6170$-$6200 & 7050 & $>8040$ & Y & 10.8$-$11.7 & \\
ST\,Her & {\sc hip}77619 &5710$-$5760 & 7340$-$7390 &8320$-$8520 & Y & 3.1$-$3.7 & unresolved \emph{HIPPARCOS} ``Problem star''$^{m}$\\
g\,Her & {\sc hip}80704 & 5770 & 6400 & 7230 & Y & 1.6 & unresolved \emph{HIPPARCOS} ``Problem star''$^{m}$\\
13\,Sge & {\sc hip}98438 & 7590 & ... & ... & N & 23.8 & double star, separation 23\arcsec$-$113\arcsec$^{o}$\\
NU\,Pav & {\sc hip}98608 & 6000 & 7050 & 8270 & Y & 4.8 & \\
V4434\,Sgr & {\sc hip}99920 & 7080 & 7180 & 7390 & Y & 64.8 & double star, separation 30\arcsec$-$46\arcsec$^{o}$\\
$\omega$\,Cap & {\sc hip}102978 & 7020$-$7030 & ... & ... & N & 53.3$-$54.9 & \\
W\,Cyg & {\sc hip}106642 & 7340 & ... & ... & N & 45.6 & \emph{HIPPARCOS} suspected double$^{n}$\\
S\,Gru & {\sc hip}110736 &5350$-$6180 & ... & ... & N & 2.1$-$46.8 & \\
HR\,Peg & {\sc hip}113131 & 6300 & 7020 & 7780 & Y & 7.0 & unresolved \emph{HIPPARCOS} ``Problem star''$^{m}$\\
19\,Psc & {\sc hip}117245 &5680 & 6450 & 7690 & Y & 7.7 & unresolved \emph{HIPPARCOS} ``Problem star''$^{m}$ \\
\hline
\end{tabular}
\end{table*}

\subsection{A criterium for binarity}
\label{criterium}

One of the objectives of this paper is to define a procedure for
selecting potential binary candidates.
Table \ref{table-famaey} lists 12 confirmed binary AGB stars,
discovered after the monitoring of their radial velocities
\citep{Famaey09}.
All of them have been detected in the near-UV and all of them, 
but {\sc hd}108907, have also a far-UV counterpart.
This is especially significant because whilst $Q_{\it NUV}$ varies
between $10 \sim 200$, the excess in the far-UV band is much larger,
$Q_{\it FUV}> 2 \times 10^5$.
This means that, as the near-UV flux has relative contributions
of the primary and the secondary, the far-UV flux is originated
mostly in the secondary.  
Therefore it can be proposed that the detection of an AGB star in
the \emph{GALEX} far-UV band is probe of binarity.

A second criterium to select binary AGB candidates is
the near-UV excess, expressed as $Q_{\it NUV}$.
Although disentangling the various effects that contribute for the
flux variability (and as a consequence $Q_{\it NUV}$) would require
a modelization of each individual binary system, it is possible to
establish a general statistical criterium for the majority
of the objects. A $U$-band amplitude of 3 mag
(a factor 16 in flux) has been observed in the light curve of several
types of variable AGB stars \citep{Celis86}, which suggests that
near-UV excesses in the approximate interval $Q_{\it NUV} < 16$
can be partially explained by the AGB variability.
Accounting for this intrinsic AGB variability, we propose $Q_{\it NUV}>20$
as an additional criterium for selecting binary AGB candidates.
The lower limit of this quotient is uncertain, and in any case
binarity should be confirmed only after a careful monitoring of
the radial velocity.

The assumption of far-UV detection {\it and/or} $Q_{\it NUV} > 20$ as a
general criterium to select binary AGB stars would select the following AGB
stars of this study as binaries: all the objects studied by
\citet{Sahai08,Sahai11} shown in Table \ref{table-sahai}; all the
AGB stars selected from \citet{Famaey09} listed in Table \ref{table-famaey};
34 objects among the 53 listed in Table \ref{table-Ita}.
These are all listed in Table \ref{table-dobles}, as well as the $T_{\it NUV}$,
$T_{\it FUV}$ and $T_{\it NUV/FUV}$ companion temperatures determined with the
methods described in Sect. \ref{temperature}.
The incidence of binary candidates in a sample limited to the distance of
500 pc can be estimated as follows.
As explained in Sect.~\ref{results-nearby}, among the 58 AGB sources
surveyed by \emph{GALEX} not known to be symbiotic systems, 5 were
not detected in the survey and 53 were detected in the far- and/or
near-UV bands.
Among these, 34 stars have been selected as binary AGB candidates.
Therefore, the occurrence of candidates is 34/58 or 59\%, higher than
some previous studies focused on other samples: 8\% of spectroscopic
binaries in field red giants \citep{Jorissen04} and 26\% in three open
clusters \citep{Mermilliod01}.
Additional work, e.g.\ based on kinematic data, should be devoted to confirm
the candidates selected in the present study, which eventually may decrease
the incidence of confirmed AGB binary systems.

\section{Conclusions}

This work has focused on the search for main-sequence (MS) companions
of AGB stars detected in the {\sc galex} survey.
This study extends the method originally proposed by \citet{Sahai08}:
the UV excess of AGB stars is due to the presence of a companion, which
can be detected above the emission of the primary if its temperature is
$T_{\rm eff} >$ 6000~K and the flux is within the detection limits of the
{\sc galex} survey.
After analysing the characteristics of the UV emission of a sample of
{\it bona-fide} binary AGB stars detected with {\sc galex}, it can be
concluded that:

\begin{enumerate}

\item Within a distance of 500 pc,
{\sc galex} detection limits of MS stars are $T_{\rm eff} >$5000~K and
$>$6500~K for the near-UV and far-UV bands, respectively.

\item Although the UV flux originates mostly from the secondary star,
other extrinsic factors might affect its intensity,
causing variability, for instance, the intrinsic variability of the primary
and the opacity of a common envelope of gas and dust
or accretion of material onto the secondary or onto an accretion disk
around it.

\item The {\sc galex} near-UV excess of confirmed binary AGB
stars varied within $10 < Q_{\it NUV} < 200$ and $Q_{\it FUV} > 2 \times 10^5$.

\item A possible criterium for binarity is: (1) detection in the far-UV
{\sc galex} band {\it and/or} (2) $Q_{\it NUV} > 20$.

\item The UV excess of the candidates to binary AGB stars cannot be fitted
with a single-temperature companion.
The UV emission of the secondary is possibly affected by the extended
atmosphere of the primary and/or its circumstellar envelope
or it may reveal additional sources of UV emission in accretion processes.

\end{enumerate}

\section*{Acknowledgements}

RO thanks the S\~ao Paulo Research Foundation (FAPESP) for the
grants \#2010/18835-3 and \#2015/00890-1.
MAG acknowledges support of the grant AYA 2014-57280-P,
co-funded with FEDER funds.  

This research has made use of the SIMBAD database, operated at CDS,
Strasbourg, France. This research has made use of NASA's Astrophysics
Data System.
The Guide Star Catalogue-II is a joint project of the Space Telescope
Science Institute and the Osservatorio Astronomico di Torino.
Space Telescope Science Institute is operated by the Association of
Universities for Research in Astronomy, for the National Aeronautics
and Space Administration under contract NAS5-26555.
The participation of the Osservatorio Astronomico di Torino is supported
by the Italian Council for Research in Astronomy.
Additional support is provided by European Southern Observatory, Space
Telescope European Coordinating Facility, the International GEMINI
project and the European Space Agency Astrophysics Division.
This publication makes use of data products from the Two Micron All Sky
Survey, which is a joint project of the University of Massachusetts and
the Infrared Processing and Analysis Center/California Institute of
Technology, funded by the National Aeronautics and Space Administration
and the National Science Foundation.




\begin{thebibliography}{99}

\bibitem[\protect\citeauthoryear{Ake \& Johnson}{1988}]{AJ88}Ake, T.B. \& Johnson, H.R., 1988, ApJ 327, 214

\bibitem[\protect\citeauthoryear{Baranne et al.}{1979}]{Baranne79}Baranne, A., Mayor, M. \& Poncet, J.L., 1979, Vistas in Astronomy 23, 279

\bibitem[\protect\citeauthoryear{Belczynski et al.}{2000}]{Belczynski00}Belckzynski, K., Mikilajewska, J., Munari, U., Ivison, R.J. \& Friedjung, M., 2000, A\&AS 146, 407

\bibitem[\protect\citeauthoryear{Beuzit et al.}{2008}]{Beuzit08}Beuzit, J.-L., Feldt, M., Dohlen, K., Mouillet, D., Puget, P., Wildi, F., Abe, L., Antichi, J., Baruffolo, A., Baudoz, P. et al., 2008, SPIE 7014, 18

\bibitem[\protect\citeauthoryear{Bianchi et al.}{2007}]{Bianchi07}Bianchi, L., Rodriguez-Merino, L., Viton, M., Laget, M., Efremova, B., Herald, J., Conti, A., Shiao, B., Gil de Paz, A., Salim, S. et al., 2007, ApJS 173, 659

\bibitem[\protect\citeauthoryear{Castelaz \& McCollum}{1995}]{CM95}Castelaz, M.W. \& McCollum, B., 1995, AJ 109, 341

\bibitem[\protect\citeauthoryear{Castelli \& Kurucz}{2003}]{CK03}Castelli, F. \& Kurucz, R.L., 2003, in IAU Symp. 210, Modelling of Stellar Atmospheres, ed. W.W. Piskunov \& D.F. Gray (San Francisco: ASP), A20

\bibitem[\protect\citeauthoryear{Celis}{1986}]{Celis86}
Celis, S.L., 1986, ApJS 60, 879

\bibitem[\protect\citeauthoryear{Corradi \& Schwarz}{1993}]{CS93}
Corradi, R.L.M. \& Schwarz, H.E., 1993, A\&A 268, 714

\bibitem[\protect\citeauthoryear{Cotton et al.}{2010}]{Cotton10}
Cotton, W.D., Ragland, S., Pluzhnik, E.A., Danchi, W.C., Traub, W.A.,
Willson, L.A. \& Lacasse, M.G., 2010, ApJS 188, 506

\bibitem[\protect\citeauthoryear{Cox}{2002}]{Cox02}
Cox, A.N. 2002, {\it Allen's Astrophysical Quantities}, Springer

\bibitem[\protect\citeauthoryear{Cox et al.}{1012}]{Cox12}Cox, N.L.J., Kerschbaum, F., van Marle, A.-J., Decin, L., Ladjal, D., Mayer, A., Groenewegen, M.A.T., van Eck, S., Royer, P., Ottensamer, R. et al., A\&A 537, A35

\bibitem[\protect\citeauthoryear{Danilovich et al.}{2015}]{Danilovich15}Danilovich, T., Olofsson, G., Black, J.H., Justtanont, K., \& Olofsson, H., 2015, A\&A 574, A23

\bibitem[\protect\citeauthoryear{De Beck et al.}{2010}]{Beck10}De Beck, E., Decin, L, de Koter, A., Justtanont, K., Verhoelst, T., Kemper, F. \& Menten, K.M., 2010, A\&A 523, 18

\bibitem[\protect\citeauthoryear{De Marco et al.}{2013}]{DeMarco13}De Marco, O., Passy, J.-C., Frew, D.J., Moe, M. \& Jacoby, G.H., 2013, MNRAS 428, 2118


\bibitem[\protect\citeauthoryear{ESA}{1997}]{ESA97}ESA 1997, The Hipparcos and Tycho Catalogues, SP 1136 (ESA)

\bibitem[\protect\citeauthoryear{Famaey et al.}{2009}]{Famaey09}Famaey, B., Pourbaix, D., Frankowski, A., van Eck, S., Mayor, M., Udry, S. \& Jorissen, A., 2009, A\&A 498, 627

\bibitem[\protect\citeauthoryear{Fluks et al.}{1994}]{Fluks94}Fluks, M.A., Plez, B., Th\'e, P.S., de Winter, D., Westerlund, B.E. \& Steenman, H.C., 1994, A\&AS 105, 311

\bibitem[\protect\citeauthoryear{Frankowski et al.}{2007}]{Fran07}Frankowski, A., Jancart, S. \& Jorissen, A., 2007, A\&A 464, 377

\bibitem[\protect\citeauthoryear{Froebrich et al.}{2010}]{Froebrich10}Froebrich, D., Schmeja, S., Samuel, D. \& Lucas, P.W., 2010, MNRAS 409, 1281

\bibitem[\protect\citeauthoryear{Fusco et al.}{2014}]{Fusco14}Fusco, T., Sauvage, J.-F., Petit, C., Costille, A., Dohlen, K., Mouillet, D., Beuzit, J.-L., Kasper, M., Suarez, M., Soenke, C. et al., 2014, SPIE 9148, 1

\bibitem[\protect\citeauthoryear{Garc\'\i a-Segura et al.}{1999}]{GS99}Garc\'\i a-Segura, G., Langer, N., R\'ozyczka, M. \& Franco, J., 1999, ApJ 517, 767

\bibitem[\protect\citeauthoryear{Garc\'\i a-Segura et al.}{2005}]{GS05}Garc\'\i a-Segura, G., L\'opez, J.A. \& Franco, J., 2005, ApJ 618, 919

\bibitem[\protect\citeauthoryear{Glass et al.}{1995}]{Glass95}Glass, I.S., Whitelock, P.A., Catchpole, R.M. \& Feast, M.W., 1995, MNRAS 273, 383

\bibitem[\protect\citeauthoryear{Groenewegen et al.}{1992}]{Groenewegen92}Groenewegen, M.A.T., de Jong, T., van der Bliek, N.S., Slijkhuis, S. \& Willems, F.J., 1992, A\&A 253, 150

\bibitem[\protect\citeauthoryear{Groenewegen \& Whitelock}{1996}]{GW96}Groenewegen, M.A.T. \& Whitelock, P.A., 1996, MNRAS 281, 1347

\bibitem[\protect\citeauthoryear{Horch et al.}{2011}]{Horch11}Horch, E.P., Gomez, S.C., Sherry, W.H., Howell, S.B., Ciardi, D.R., Anderson, L.M. \& van Altena, W.F., 2011, AJ 141, 45

\bibitem[\protect\citeauthoryear{Houk et al.}{1988}]{Houk88}Houk, M. \& Smith-Moore, M., 1988, {\it Michigan Spectral Survey}, Univ. Michigan

\bibitem[\protect\citeauthoryear{Huggins et al.}{2009}]{Huggins09}Huggins, P.J., Mauron, N. \& Wirth, E.A., 2009, MNRAS 396, 1805

\bibitem[\protect\citeauthoryear{Hunsch et al.}{1998}]{Hunsch98}Hunsch, M., Schmitt, J.H.M.M., Schroeder, K. \& Zickgraf, F., 1998, A\&A 330, 225

\bibitem[\protect\citeauthoryear{Ita et al.}{2010}]{Ita10}Ita, Y., Matsuura, M.,
Ishihara, D., Oyabu, S., Takita, S., Kataza, H., Yamamura, I., Matsunaga, N.,
Tanab\'e, T., Nakada, Y. et al., 2010, A\&A 514, A2

\bibitem[\protect\citeauthoryear{Ivanova et al.}{2013}]{Ivanova13}Ivanova N., et al., 2013, A\&ARv, 21, 59 

\bibitem[\protect\citeauthoryear{Johnson et al.}{1993}]{JAA93}Johnson, H.R., Ake, T.B. \& Ameen, M.M., 1993, ApJ 402, 667

\bibitem[\protect\citeauthoryear{Jorissen et al.}{2004}]{Jorissen04}Jorissen, A., Famaey, B., Dedecker, M., Pourbaix, D., Mayor, M. \& Udry, S., 2004, RMxAC 21, 71

\bibitem[\protect\citeauthoryear{Jura}{1994}]{Jura94}Jura, M., ApJ 422, 102

\bibitem[\protect\citeauthoryear{Karovska et al.}{1993}]{Karovska93}Karovska, M., Nisenson, P. \& Beletic, J., 1993, ApJ 402, 311

\bibitem[\protect\citeauthoryear{Karovska et al.}{1997}]{Karovska97}Karovska, M., M., Hack, W., Raymond, J. \& Guinan, E., 1997, ApJ 482, L175

\bibitem[\protect\citeauthoryear{Keenan}{1954}]{Keenan54}Keenan, P.C., 1954, ApJ 120, 484

\bibitem[\protect\citeauthoryear{Keenan et al.}{1974}]{Keenan74}Keenan, P.C., Garrison, R.F. \& Deutsch, A.J., 1974, ApJS 28, 271

\bibitem[\protect\citeauthoryear{Kervella et al.}{2015}]{Kervella15}Kervella, P., Montarg\`es, M., Lagadec, E., Ridgway, S.T., Haubois, X., Girard, J.H., Ohnaka, K., Perrin, G. \& Gallenne, A., 2015, A\&A 578, A77

\bibitem[\protect\citeauthoryear{Koornneef}{1983}]{Koornneef83}Koornneef, J., 1983, A\&A 128, 84

\bibitem[\protect\citeauthoryear{Lasker et al.}{2008}]{Lasker08}Lasker, B.M.,
Lattanzi, M.G., McLean, B.J., Bucciarelli, B., Drimmel, R., Garcia, J.,
Greene, G., Guglielmetti, F., Hanley, C., Hawkins, G. et al., 2008, AJ 136, 735

\bibitem[\protect\citeauthoryear{Lejeune et al.}{1997}]{Lejeune97}Lejeune, T., Cuisinier, F. \& Buser, R., 1997, A\&AS 125, 229

\bibitem[\protect\citeauthoryear{L\'epine et al.}{1995}]{LOE95}L\'epine, J.R.D., Ortiz, R. \& Epchtein, N., 1995, A\&A 299, 453

\bibitem[\protect\citeauthoryear{Maercker et al.}{2014}]{Maercker14}Maercker, M., Ramstedt, S., Leal-Ferreira, M.L., Olofsson, G., \& Floren, H.G., 2014, A\&A 570, A101

\bibitem[\protect\citeauthoryear{Maercker et al.}{2016}]{Maercker16}Maercker, M., Vlemmings, W.H.T., Brunner, M., De Beck, E., Humphreys, E.M., Kerschbaum, F., Lindqvist, M., Olofsson, H. \& Ramstedt, S., 2016, A\&A 586, A5

\bibitem[\protect\citeauthoryear{Ma\'\i z-Apell\'aniz}{2006}]{MA06}Ma\'\i z-Apell\'aniz, J., 2006, AJ 131, 1184

\bibitem[\protect\citeauthoryear{Marshall et al.}{2006}]{Marshall06}Marshall, D.J., Robin, A.C., Reyl\'e, C., Schultheis, M. \& Picaud, S., 2006, A\&A 453, 635

\bibitem[\protect\citeauthoryear{Martin et al.}{2005}]{Martin05}Martin, D.C., Fanson, J., Schiminovich, D., Morrissey, P., Friedman, P.G., Barlow, T.A., Conrow, T., Grange, R., Jelinsky, P.N., Milliard, B. et al., 2005, ApJ 619, L1

\bibitem[\protect\citeauthoryear{Mason et al.}{1999}]{Mason99}Mason, B.D., Martin, C., Hartkopf, W.I., Barry, D.J., Germain, M.E., Douglass, G.G., Worley, C.E. \& Wycoff, G.L., 1999, AJ 117, 1890

\bibitem[\protect\citeauthoryear{Mason et al.}{2001}]{Mason01}Mason, B.D., Wycoff, G.L., Hartkopf, W.I., Douglass, G.G. \& Worley, C.E., 2001, AJ 122, 3466

\bibitem[\protect\citeauthoryear{Mauron et al.}{2013}]{Mauron13}Mauron, N., Huggins, P.J. \& Cheung, C.-L., 2013, A\&A 551, A110

\bibitem[\protect\citeauthoryear{Mayer et al.}{2013}]{Mayer13}Mayer, A., Jorissen, A., Kerschbaum, F., Ottensamer, R., Nowotny, W., Cox, N.L.J., Aringer, B., Blommaert, J.A.D.L., Decin, L., van Eck, S., Gail, H.-P., Groenewegen, M.A.T., Kornfeld, K., Mecina, M., Posch, Th., Vandenbussche, B. \& Waelkens, C., 2013, A\&A 549, A69

\bibitem[\protect\citeauthoryear{McDonald et al.}{2012}]{McDonald12}McDonald, I., Zijlstra, A.A. \& Boyer, M.L., 2012, MNRAS 427, 343

\bibitem[\protect\citeauthoryear{Mennessier et al.}{1997}]{Mennessier97}Mennessier, M.O., Boughaleb, H. \& Mattei, J.A., 1997, A\&AS 124, 143

\bibitem[\protect\citeauthoryear{Mermilliod}{2001}]{Mermilliod01}Mermilliod, J.-C., 2001, in Astrophys. Space Sc. Lib., 264,
The Influence of Binaries on Stellar Population Studies, ed. D. Vanbeveren
(Dordrecht: Kluwer), 3


\bibitem[\protect\citeauthoryear{Morrissey et al.}{2005}]{Morrissey05}Morrissey, P., Schiminovich, D., Barlow, T.A., Martin, D.C., Blakkolb, B., Conrow, T., Cooke, B., Erickson, K., Fanson, J., Friedman, P.G. et al., 2005, ApJ 619, L10

\bibitem[\protect\citeauthoryear{Nhung et al.}{2015}]{Nhung15}Nhung, P.T., Thi HOai, D., Winters, J.M., Darriulat, P., G\'erard, E. \& Le Bertre, T., 2015, RAA 15, 713

\bibitem[\protect\citeauthoryear{Nordhaus et al.}{2007}]{Nordhaus07}Nordhaus, J., Blackman, E.G. \& Frank, A., 2007, MNRAS 376, 599

\bibitem[\protect\citeauthoryear{Ortiz \& L\'epine}{1993}]{OL93}Ortiz, R. \& L\'epine, J.R.D., 1993, A\&A 279, 90

\bibitem[\protect\citeauthoryear{Ortiz \& Maciel}{1996}]{OM96}Ortiz, R. \& Maciel, W.J., 1996, A\&A 313, 180

\bibitem[\protect\citeauthoryear{Perryman et al.}{1997}]{Perryman97}Perryman, M.A.C., Lindegren, L., Kovalevsky, J., Hog, E., Bastian, U., Bernacca, P.L., Cr\'ez\'e, M., Donati, F., Grenon, M., Grewing, M. et al., 1997, A\&A 323, L49

\bibitem[\protect\citeauthoryear{Pourbaix et al.}{2003}]{Pourbaix03}Pourbaix, D., Platais, I., Detournay, S., Jorissen, A., Knapp, G. \& Makarov, V.V., 2003, A\&A 399, 1167

\bibitem[\protect\citeauthoryear{Pourbaix et al.}{2004}]{Pourbaix04}Pourbaix, D., Tokovinin, A.A., Batten, A.H. et al. 2004, A\&A 424, 727

\bibitem[\protect\citeauthoryear{Prieur et al.}{2002}]{Prieur02}Prieur, J.L., Aristidi, E., Lopez, B., Scardia, M., Mignard, F. \& Carbillet, M., 2002, ApJS 139, 249

\bibitem[\protect\citeauthoryear{Ramstedt et al.}{2011}]{Ramstedt11}Ramstedt, S., Maercker, M., Olofsson, G., Olofsson, H., Schoeier, F.L., 2011, 531, A148

\bibitem[\protect\citeauthoryear{Sahai et al.}{2008}]{Sahai08}Sahai, R., Findeisen, K., Gil de Paz, A \& S\'anchez-Contreras, C., 2008, ApJ 689, 1274

\bibitem[\protect\citeauthoryear{Sahai et al.}{2011}]{Sahai11}Sahai, R., Neill, J.D., Gil de Paz, A. \& S\'anchez-Contreras, C., 2011, ApJ 740, L39

\bibitem[\protect\citeauthoryear{Sahai et al.}{2015}]{Setal15}
Sahai R., Sanz-Forcada J., S{\'a}nchez Contreras C., Stute M., 2015,
ApJ, 810, 77 

\bibitem[\protect\citeauthoryear{Salaris \& Cassisi}{1997}]{SC97}Salaris, M. \& Cassisi, S., 1997, MNRAS 289, 406


\bibitem[\protect\citeauthoryear{Samus et al.}{2015}]{Samus15}Samus, N.N., Durlevich, O.V., Goranskij, V.P., Kazarovets, E.V., Kireeva, N.N., Pastukhova, E.N., Zharova, A.V., 2007-2015, {\it General Catalogue of Variable Stars} online edition available at www.sai.msu.su/gcvs/gcvs

\bibitem[\protect\citeauthoryear{Sanchez et al.}{2015}]{Sanchez15}Sanchez, E., Montez Jr., R., Ramstedt, S. \& Stassun, K.G., 2015, ApJ 798, L39

\bibitem[\protect\citeauthoryear{Sivagnanam et al.}{1988}]{Sivagnanam88}Sivagnanam, P., Le Squeren, A.M. \& Foy, F., 1988, A\&A 206, 285

\bibitem[\protect\citeauthoryear{Skrutskie et al.}{2006}]{Skrutskie06}Skrutskie, R.M., Cutri, R. Stiening, M.D. Weinberg, S. Schneider, J.M. Carpenter, C. Beichman, R. Capps, T. Chester, J. Elias, J. Huchra, J. Liebert, C. Lonsdale, D.G. Monet, S. Price, P. Seitzer, T. Jarrett, J.D. Kirkpatrick, J. Gizis, E. Howard, T. Evans, J. Fowler, L. Fullmer, R. Hurt, R. Light, E.L. Kopan, K.A. Marsh, H.L. McCallon, R. Tam, S. Van Dyk, and S. Wheelock, 2006, AJ, 131, 1163

\bibitem[\protect\citeauthoryear{Sloan \& Price}{1998}]{SP98}Sloan, G.C. \& Price, S.D., 1998, ApJS 119, 141

\bibitem[\protect\citeauthoryear{Smith \& Lambert}{1987}]{SL87}Smith, V.V. \& Lambert, D.L., 1987, AJ 94, 977

\bibitem[\protect\citeauthoryear{Soker}{1998}]{Soker98}Soker, N., 1998, ApJ 496, 833

\bibitem[\protect\citeauthoryear{Soker}{2006}]{Soker06}Soker, N., 2006, PASP 118, 260

\bibitem[\protect\citeauthoryear{Staff et al.}{2016}]{Staff16}Staff J.~E., De Marco O., Macdonald D., Galaviz P., Passy J.-C., Iaconi R., Low M.-M.~M., 2016, MNRAS, 455, 3511 

\bibitem[\protect\citeauthoryear{Thiering \& Reimers}{1993}]{TR93}Thiering, I. \& Reimers, D., 1993, A\&A 274, 838

\bibitem[\protect\citeauthoryear{van Langevelde, H.J. et al.}{1990}]{vL90}van Langevelde, H.J., van der Heiden, R. \& Schooneveld, C., 1990, A\&AS 239, 193

\bibitem[\protect\citeauthoryear{van Leeuwen}{2007}]{Leeuwen07}van Leeuwen, F., 2007, A\&A 474, 653

\bibitem[\protect\citeauthoryear{Wood \& Cahn}{1977}]{WC77}Wood, P.R. \& Cahn, J.H., 1977, ApJ 211, 499

\bibitem[\protect\citeauthoryear{Yuan et al.}{2013}]{Yuan13}Yuan, H.B., Liu, X.W. \& Xiang, M.S., 2013, MNRAS 430, 2188

\end{thebibliography}


\bsp	
\label{lastpage}
\end{document}